\documentclass[twocolumn,aps,prb,showpacs]{revtex4}
\usepackage{graphicx}

\begin{document}

%
%

\title{Electron-elastic-wave interaction in a two-dimensional topological insulator}

\author{Wu Xiao-Guang (ÎâÏþ¹â)}

\affiliation{SKLSM, Institute of Semiconductors, Chinese Academy
of Sciences, Beijing 100083, China}

\begin{abstract}

The interaction between an electron and an elastic wave is investigated
for HgTe and InAs-GaSb quantum wells.  The well-known Bernevig-Hughes-Zhang model,
i.e., the $4\times 4$ model for a two-dimensional (2D) topological insulator (TI),
is extended to include terms that describe the coupling between the
electron and the elastic wave.  The influence of this interaction on the
transport properties of the 2DTI and of the edge states is discussed.
As the electron-like and hole-like carriers interact with the elastic wave
differently due to the cubic symmetry of the 2DTI, one may utilize the elastic
wave to tune/control the transport property of charge carriers in the 2DTI.
The extended 2DTI model also provides the possibility to investigate the
backscattering of edge states of a 2DTI more realistically.

\end{abstract}


\maketitle

%
%

When a two-dimensional (2D) topological insulator (TI) has a boundary
with a normal insulator or vacuum, it is predicted theoretically that
there exist gapless edge states.\cite{rmp1,rmp2,bhz}  The existence of
such edge states have been confirmed experimentally.\cite{exp1,exp2,exp3,exp4}
When the system is time reversal invariant, these edge states will not
be affected by an elastic back scattering center.\cite{rmp1,rmp2,bhz}
Thus, one has a transport channel for the charge carriers, i.e., the
information carriers, with no energy dissipation, and this will be of
great application importance in the information technology.

The edge states resulted from the 2DTI have been intensively studied
both theoretically and experimentally in recent years.  There are
studies on the phonon induced back scattering in helical edge states,\cite{epi}
the inelastic back scattering due to electron-electron interaction,\cite{eei}
the influence of a charge puddle,\cite{puddle} and the interaction
between edge states and nuclear spins.\cite{rkky}  There are also
investigations concerning the probing and tuning of transport property
of edge states, e.g., via a quantum point contact and via an artificially
implemented charge puddle.\cite{zhang,fabian,pud}

In this paper, we propose another mechanism to tune/control the transport
property of edge states in a 2DTI.  The idea behind this mechanism is simple.
A 2DTI is embedded in a solid made of a crystal structure, and this solid
can support an elastic wave.\cite{kittel,landau}  The elastic wave can change
the strain in the solid and thus can affect the electronic property.
In fact, many TI materials are obtained by carefully manipulating the
static strain in the material.\cite{strain}  It is also possible to tune
the strain dynamically.\cite{phononics}  In this paper, we focus on
the 2DTI system realized in a HgTe quantum well and in an InAs-GaSb
quantum well.  The quantum wells considered have the [001] axis as the
growth direction.

A 2DTI system can be described by the well-known Bernevig-Hughes-Zhang
model.\cite{bhz}  It is a $4\times 4$ Hamiltonian and it reads
$$ H({\bf k}) =
   \varepsilon(k) I +
   \left( \matrix{
   h({\bf k}) & 0 \cr
   0 & h^*(-{\bf k}) \cr
   } \right)
\phantom{...}, $$
with
$$ h({\bf k}) =
   \left( \matrix{
   M(k) & -A(k_x+ik_y) \cr
   -A(k_x-ik_y) & -M(k) \cr
   } \right)
\phantom{...}, $$
and $\varepsilon(k)=C-Dk^2$ and $M(k)=M-Bk^2$.  ${\bf k}$ is the
in-plane wave vector of the electron.
In this Hamiltonian, the upper and lower $2\times 2$ spin blocks
are not coupled.

When a symmetric quantum well is biased by an externally applied gate
voltage along the growth direction, a coupling between the upper
block and lower spin block is introduced.  This coupling adds the following
term to the Hamiltonian
$$ H_{1}({\bf k}) =
   \left( \matrix{
   0 & h_1({\bf k}) \cr
   h_1^\dagger({\bf k}) & 0 \cr
   } \right)
\phantom{...}, $$
with
$$ h_1({\bf k}) =
   \left( \matrix{
   A_1(k_y + i k_x) & 0 \cr
   0 & 0 \cr
   } \right)
\phantom{...}. $$
For a $8$ nm HgTe quantum well, one finds
that $A_1=192 \times 10^{-5} E_z$ meV nm.  The bias electric field $E_z$
is in the unit of V/cm.  When the quantum well is asymmetric, like the
case of an InAs-GaSb quantum well, the value of $A_1$ is determined by
the specific structure parameters of the quantum well.

In a cubic crystal like HgTe, InAs, and GaSb, an externally applied
elastic wave or strain potential is characterized by six quantities
$u_{\alpha,\beta}=(\partial u_{\alpha}/\partial x_{\beta}
+\partial u_{\beta}/\partial x_{\alpha})/2$ with $\alpha$ and $\beta$
stand for $x$, $y$, $z$, respectively.\cite{kittel,landau}
Note that, $u_{\alpha,\beta}$ can be spatially and temporally dependent.
The $x$-axis is chosen along the [100] direction,
and the $y$-axis is along the [010] direction.
Within the elastic deformation potential approximation and the adiabatic
approximation, this strain introduces an additional Hamiltonian
describing the interaction between the electron and the elastic wave
\begin{equation}
   H_{int} =
   \left( \matrix{
   h_{11} & h_{12} & 0 & h_{14} \cr
   h^\dagger_{12} & h_{22} & -h_{14} & 0 \cr
   0 & -h^\dagger_{14} & h_{11} & h^\dagger_{12} \cr
   h^\dagger_{14} & 0 & h_{12} & h_{22} \cr
   } \right)
   \phantom{...},
\end{equation}
where
\begin{eqnarray*}
&& h_{11} = a_{11}( u_{xx} + u_{yy} ) + b_{11} u_{zz}
  \phantom{...}, \\
&& h_{22} = a_{22}( u_{xx} + u_{yy} ) + b_{22} u_{zz}
  \phantom{...}, \\
&& h_{12} = a_{12}( u_{yz} - i u_{xz} )
  \phantom{...}, \\
&& h_{14} = i a_{14}( u_{yy} - u_{xx} ) + b_{14} u_{xy}
  \phantom{...}. \\
\end{eqnarray*}
This is the central result of this paper.
In Eq.(1), $h_{11}$ describes
the interaction between the electron-like component and the
elastic wave.  $h_{22}$ is for the hole-like component.  $h_{12}$
couples the electron-like and hole-like components.  $h_{14}$
couples directly two spin blocks.  Note that, $u_{\alpha,\beta}$
enters the above $h_{ij}$ terms in different ways.  This is because
of the cubic symmetry of the crystal structure considered.\cite{kittel,landau}

One can generate elastic waves with various characteristics, e.g.,
a longitudinal wave, a shear wave, and various propagating directions.\cite{kittel,landau}
Then, one can utilize such designed elastic waves to manipulate the transport
properties in a 2DTI and of the edge states.  In the absence of $h_{14}$,
the elastic wave can be used to drive the charge carrier in the 2DTI
like an ocean tide wave drives a surfer.  This can be viewed as a charge
pumping process, with the same pumping effect for both spin blocks.
When $h_{14}$ becomes nonzero, the elastic waves applied not only drive
the charge carriers in each spin blocks, but also provide a channel
for charge carriers transferring across two spin blocks.  One can take
the 2DTI model with the zero wave function boundary condition, and deduce
a low energy model for the edge states.  Due to the cubic symmetry of
the solid considered, it is evident that, the interaction between the
edge states and the elastic wave will not just be proportional to $u_{xx}+u_{yy}
+u_{zz}$.  This can be easily understood, as the states in the
2DTI and in the 2DTI edge states are mixed states originated
from the conduction band and valence band electronic states.
This provides one a possibility to tune the coupling of edge
states with the elastic wave.  A more detailed
investigation on the influence of the electron elastic wave interaction
in the 2DTI and on the transport property of edge states will be reported
elsewhere.  It is interesting to note that the elastic wave can be
utilized to pump electrons in quantum dots, to transport optically
generated excitons, and to route quantum information between quantum
bits.\cite{saw1,saw2,saw3}

Next, the values of parameters $a_{ij}$ and $b_{ij}$ appearing
in $h_{ij}$ are provided for specific HgTe and InAs-GaSb quantum well
structures, respectively.
For a HgTe quantum well with well width $8$ nm, it is found that
\begin{eqnarray*}
 && a_{11} = -0.63 {\rm eV}
    \phantom{...}, \phantom{...}
    b_{11} = -1.85 {\rm eV}
    \phantom{...}, \\
 && a_{22} = 0.19 {\rm eV}
    \phantom{...}, \phantom{...}
    b_{22} = 2.80 {\rm eV}
    \phantom{...}. \\
\end{eqnarray*}
They are almost independent from the bias voltage, when a bias is
applied along the growth direction of the symmetric quantum well.
It is found that
\begin{eqnarray*}
 && a_{12} = 1.3\times 10^{-5} E_z {\rm eV}
    \phantom{...}, \\
 && a_{14} = 0.6\times 10^{-5} E_z {\rm eV}
    \phantom{...}, \\
 && b_{14} = -1.3\times 10^{-5} E_z {\rm eV}
    \phantom{...}. \\
\end{eqnarray*}
They linearly depend on the bias electric field $E_z$ which
is in the unit of V/cm.  The HgTe quantum well is enclosed by
two Hg$_{x}$Cd$_{1-x}$Te barriers, and the
wave function only penetrates slightly into the barriers.
The corresponding $a_{ij}$ and $b_{ij}$ parameters for the barriers
are rather small and can be safely neglected.

For an InAs-GaSb quantum well, the following parameters are
provided as an example.
The quantum well structure consists of a $12.5$ nm InAs layer and a $5$
nm GaSb layer.  The $H_{int}$ will have two parts, one for the InAs layer
and another one for the GaSb layer.
For the InAs layer, one has
\begin{eqnarray*}
 && a_{11} = -3.03
    \phantom{...}, \phantom{...}
    b_{11} = -3.64
    \phantom{...}, \\
 && a_{22} = 0
    \phantom{...}, \phantom{...}
    b_{22} = 0.02
    \phantom{...}, \\
 && a_{12} = 0.04
    \phantom{...}, \\
 && a_{14} = 0.01
    \phantom{...}, \phantom{...}
    b_{14} = -0.03
    \phantom{...}. \\
\end{eqnarray*}
For the GaSb layer, one has
\begin{eqnarray*}
 && a_{11} = 0.28
    \phantom{...}, \phantom{...}
    b_{11} = -0.27
    \phantom{...}, \\
 && a_{22} = -0.45
    \phantom{...}, \phantom{...}
    b_{22} = 2.74
    \phantom{...}, \\
 && a_{12} = 2.11
    \phantom{...}, \\
 && a_{14} = 0.85
    \phantom{...}, \phantom{...}
    b_{14} = -2.32
    \phantom{...}. \\
\end{eqnarray*}
They are in the unit of eV.
The InAs-GaSb quantum well is enclosed by two AlSb barriers, and the
wave function only penetrates slightly into the barriers, though the
the wave function from the InAs layer and GaSb layer are strongly
hybridized.  The contribution from the barriers is small and can be
safely neglected.

Next, we describe briefly the derivation of Eq.(1) and the evaluation
of related parameters.  The calculation of one-electron energy levels
of HgTe and InAs-GaSb quantum wells
is based on the well documented eight-band ${\bf k}\cdot{\bf p}$ approach.
For details about this method, e.g., the operator ordering, the influence
of remote bands, the influence of strain, and the modification due to
hetero-junction interfaces, we refer to a partial list of publications
and references therein.\cite{winkler,bahder,burt,foreman,smith,zawadzki,rossler,chao}

The quantum well is assumed to be parallel to the $xy$ plane, and
the $z$ direction is along the growth direction of the quantum well.
The HgTe quantum well structure consists of a left and a right
Cd$_{x}$Hg$_{1-x}$Te barriers.  Enclosed by the barriers is the HgTe
quantum well, and $x=0.3$ is used in the calculation.
The InAs-GaSb quantum well studied in this paper have the following
structure: a left AlSb barrier, the InAs layer, the GaSb layer,
and finally a right AlSb barrier.  The growth direction of the
quantum wells is assumed to be [001].
The material specific parameters, i.e., the band parameters, used
in the present calculation are widely used in the literature.\cite{bp1,bp2,bp3,bp4}

When an elastic wave is applied to the quantum well, for each layer
one has a correction due to the strain induced by the elastic wave.
In the eight-band ${\bf k}\cdot{\bf p}$ approximation, one has the
following Hamiltonian
$$ \left( \matrix{
   s_1 & s_3 \cr
   s_3^\dagger & s_2 \cr
   } \right)
\phantom{...}, $$
with $s_1$ given by
$$ \left( \matrix{
   a'e & \sqrt{2}w & w & 0 & t^\dagger & \sqrt{2}t^\dagger \cr
     & p+q & \sqrt{2}q & t^\dagger & 0 & \sqrt{3/2}s^\dagger \cr
     & & ae & -\sqrt{2}t^\dagger & \sqrt{3/2}s^\dagger & 0 \cr
     & & & a'e & \sqrt{2}w & w^\dagger \cr
     & & & & p+q & -\sqrt{2}q \cr
     & & & & & ae \cr
   } \right)
\phantom{...}, $$
$$ s_2 =
   \left( \matrix{
   p-q & 0 \cr
   & p-q \cr
   } \right)
\phantom{...}, $$
$$ s_3 =
   \left( \matrix{
   0 & -\sqrt{3}t \cr
   r^\dagger & s \cr
   -\sqrt{2}r^\dagger & \sqrt{1/2}s \cr
   -\sqrt{3}t^\dagger & 0 \cr
   -s^\dagger & r \cr
   \sqrt{1/2}s^\dagger & \sqrt{2}r \cr
   } \right)
\phantom{...}. $$
This is the wave vector independent part.  There is another
contribution from the spin-orbit interaction, but it will not
be given here for simplicity.  It can be found in the
literature.\cite{winkler,bahder,burt,foreman,smith,zawadzki,rossler,chao}
Note that, in the above equation $s_1$ and $s_2$ are hermitian.

In the above $s_i$ terms, one has
\begin{eqnarray*}
&& e = u_{xx} + u_{yy} + u_{zz}
  \phantom{...}, \\
&& w = ib' u_{xy}/\sqrt{3}
  \phantom{...}, \\
&& t = b'(u_{xz} + iu_{yz})/\sqrt{6}
  \phantom{...}, \\
&& p = a (u_{xx} + u_{yy} + u_{zz})
  \phantom{...}, \\
&& q = b (u_{zz} - (u_{xx} + u_{yy})/2)
  \phantom{...}, \\
&& r = \sqrt{3}b(u_{xx} - u_{yy})/2 - id u_{xy}
  \phantom{...}, \\
&& s = -d(u_{xz} - iu_{yz})
  \phantom{...}, \\
\end{eqnarray*}
with $a'$, $b'$, $a$, $b$, and $d$ the deformation potential parameters.
These parameters can be found in the literature for the material
considered.\cite{bp1,bp2,bp3,bp4}  One can decompose the above
Hamiltonian into ones that is proportional to $u_{ij}$.
Then, Eq.(1) is obtained by taking the matrix elements for
every $u_{ij}$ terms.  This is the same approach in obtaining
the well-known Bernevig-Hughes-Zhang model.\cite{bhz}

It should be pointed out that the interaction Hamiltonian
given by Eq.(1) also provides a more realistic model for the
study of the backscattering of edge states in the 2DTI due to
interaction between the charge carrier and the acoustic phonon.
The acoustic phonon, as an intrinsic elementary excitation of
the crystal, deforms the solid and affects the electronic
properties, especially the transport properties.\cite{kittel}
The scattering of charge carriers by phonons is an important
source of decoherence and dephasing for electrons and electron
spins.\cite{datta}

%
%

In summary, the interaction between an electron and an elastic wave
is investigated theoretically for HgTe and InAs-GaSb quantum wells.
The well-known Bernevig-Hughes-Zhang model, which is widely used to
investigate the 2DTI properties, is further extended to include terms
that describe the coupling between the electron and the elastic wave.
The influence of this interaction, which can be considered as a
tuning/controlling mechanism, on the transport properties of the 2DTI
and of the edge states is discussed.  The extended model also provides
the possibility to investigate the backscattering of edge states by
acoustic phonons in a more realistic way.

This work was partly supported NSF of China via
projects 61076092 and 61290303.

%
%

\end{document}